\documentclass[intlimits,twoside,a4paper]{article}

\usepackage[cp1251]{inputenc}

\usepackage{cmpj3}
\usepackage{graphicx}        

\newcommand{\be}{\begin{equation}}
\newcommand{\ee}{\end{equation}}
\newcommand{\bea}{\begin{eqnarray}}
\newcommand{\eea}{\end{eqnarray}}
\articletype{A part of the Special collection to the 100th anniversary of the birth of Ihor Yukhnovskii}

\issue{2026}{29}{2}{23501}
\doinumber{10.5488/CMP.29.23501}
\title[On the statistical theory of strong electrolytes and high-temperature plasmas]{On the statistical theory of strong electrolytes and high-temperature plasmas: new applications of the work of Yukhnovskii and Kelbg} 

\author[W. Ebeling, M. Holovko]{W. Ebeling\orcid{0000-0003-0740-3016}\refaddr{label1}\thanks{Corresponding author:\email{ebeling@physik.hu-berlin.de}.},  M. Holovko\orcid{0000-0001-8114-5356}\refaddr{label2}}
\addresses{
	\addr{label1}{Institute of Physics, Humboldt University, Berlin, Germany}
\addr{label2}{Yukhnovskii Institute for Condensed Matter Physics of the National Academy of Sciences of Ukraine, 1~Svientsitskii Str., 79011, Lviv, Ukraine}}

\date{Received 11 January 2026; revised 23 February 2026; accepted 31 March 2026; published 29 June 2026}

\begin{document}
	
	\maketitle
\begin{abstract}
Remembering here the work of two pioneers of the statistical physics of Coulomb systems,
G\"unter Kelbg,  and Ihor Yukhnovskii, we analyze their methods and
give some new applications
to ionic solutions and quantum plasmas. In particular, we develop applications of the theory to strong electrolytes and to thermal high-temperature plasmas at $T > 10^5$~K using the exponential interaction model.
We show the strong structural similarity of these two classes of Coulomb systems, which 
physics is determined mostly by contributions proportional to $e^4$ and~$e^6$.
We predict at higher densities a structural transition to oscillating correlations. The thermodynamic functions show a smooth transition 
from a quadratic root  increase to a slower increase like $n_i^{1/4}$ which observes the Onsager bound. 
Effects of asymmetries in charges and masses are studied  with applications to ionic systems with
multiple charges  and to high-temperature plasmas, in particular, to plasmas with He$^{2+}$-ions.
\printkeywords 
 \end{abstract}

\section{Introduction}
\label{intro}

The statistical theory of systems of charged particles was the main topic of the two old friends,
the pioneers of Coulombic systems, G\"unter~Kelbg (1922--1988) and Ihor~R.~Yukhnovskii (1925--2024) both nearly of same age. 
%
Both were born in the 1920s along nearly the same meridian of longitude: G\"{u}nter Kelbg
in K\"{o}nigsberg and Ihor Yukhnovskii in the village of Kniahynyne, within the Volyn Voivodeship.
Ihor Yukhnovskii started to work in our field already in  the early fifties when he  developed  a theory of electrolytes with  his adviser Glauberman in Lviv \cite{Yukhn02,YukhnC26,GlaYuk52,Yukhn54,Yukhn58}. This theory was  based on the method of Bogoliubov~\cite{Bogolyubov,Bogolyubov_b} and later extended to the range of higher concentrations in many subsequent works, in particular, by using collective coordinates  \cite{Yukhn02,Yukhn58,Yukhn80,Yukhn25,Falkenhagen,Falkenhagen_b,Tjablikov,Barthel}. Yukhnovskii and Glauberman used for the description of the forces between charged particles the effective potential already proposed in 1927 by Kramers in quantum theory and since 1934 by Hellman in quantum chemistry. We call this class of potentials exponential potentials and consider the simplest form 
\bea
V_{ij} (r) = \frac{e_i e_j}{4 \piup  \epsilon r} [1 - \exp(-\alpha r)], \qquad \epsilon = \epsilon_0 \epsilon_r,
\label{eq-1}
\eea
where $e_i$ are the ionic charges, $\epsilon$ is the dielectric constant, and $\alpha$ is a characteristic reciprocal length
which in more refined versions depends on the pair index $\alpha_{ij}$. We will shift here the species-dependence into the 
short-range part of the interactions. 
Glauberman and Yukhnovskii used this class of potentials  first for 
solving equilibrium problems of electrolyte theory \cite{GlaYuk52,Yukhn80,Barthel}. G\"unter Kelbg studied physics and mathematics at the University of Rostock and  in the mid-50th joined the group of Hans Falkenhagen, one
of the pioneers of electrolyte theory \cite{Falkenhagen,Falkenhagen_b,Harned}. Kelbg graduated in 1954 with a dissertation on deviations from Ohm's law in electrolytes and then turned to the equilibrium statistical thermodynamics of ionic solutions. He used 
the exponential model applying the results of Glauberman
and Yukhnovskii and developed applications to transport problems \cite{Kelbg59,Kelbg59_b}. Further, Kelbg also developed powerful concepts of collective coordinates parallel to Yukhnovskii's findings. Then Kelbg's research interest shifted to quantum effects in Coulomb systems, and in 1963/1964 there appeared his first
quantum theoretical publications in Annalen der Physik \cite{Kelbg62} with a fundamental achievement in quantum plasma theory. He derived a new effective interaction potential of Coulombic interacting
particles which is now connected with his name~\cite{Kelbg63,Kelbg63-b,BoEbal22}. He showed that the effective interaction between charges in a quantum plasma is given by a rather complicated expression, which, 
however, has a similar shape like our model exponential potential applied so far only to electrolytes \cite{Falkenhagen,Falkenhagen_b,Kelbg59,Kelbg59_b}. Kelbg extended the potential to the quantum-statistical case, estimated the parameters, and studied fruitful applications to plasmas. 
We continue here the analysis of the important contributions of Yukhnovskii and Kelbg, which we 
started already in previous works~\cite{Holovko,EbCMP25} and apologize for repeats.
The common idea of the works of G\"unter Kelbg and Ihor Yukhnovslii was to use potentials possessing a 
simple Fourier transform for an improved description of screening effects. The method permits  
to introduce collective variables and to develop a quite effective mathematical technique following methods by Bohm, Pines, and Zubarev~\cite{Yukhn54,Yukhn80,Yukhn58,Kelbg59,Kelbg59_b,Kelbg63,Kelbg63-b,KelbgHoff64,BoEbal22}.

\begin{figure}[htbp]
	\begin{center}
		\includegraphics[scale=1.3]{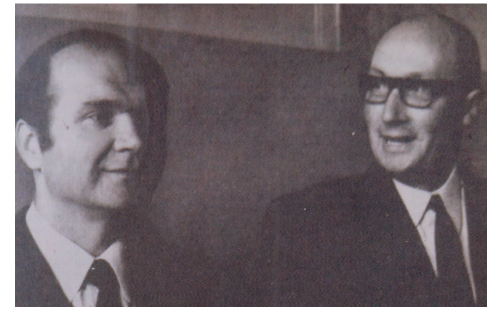}
		\caption{Ihor Yukhnovskii (left) visited Rostock in 1966 and in 1972 and got a close friend of G\"unter Kelbg (right). In reply, G\"unter Kelbg and his coworkers visited Lviv several times,
			developing a close collaboration  (photo by Dorit Hagen).}
		\label{Yukhn}
	\end{center}
\end{figure}

Further, more general models of effective interactions for electrolytes with non-additive radii
were developed  \cite{Krienke2013,EbFeKr21, EbFeKr21_b,EbKrCMP23}, in particular, in common work with Hartmut Krienke
\cite{GoKr89}. In their first approach to electrolytes, Kelbg and Yukhnovkii used the strict expansions regarding the plasma parameter which were developed in 1946 by Nikolai N. Bogoliubov, solving the so-called BBGKY hierarchy with systematic expansion methods
\cite{Bogolyubov,Bogolyubov_b}. A detailed description of the Glauberman--Yukhnovskii--Kelbg theory including an extensive representation of many successful applications to a comparison of the experimental data is given in the monograph of Falkenhagen and in other papers \cite{Yukhn80,Falkenhagen,Falkenhagen_b,Holovko,EbCMP25}.  
Despite much success in the description of data, the work on the exponential potential should still be continued. The general idea is to consider a species-independent simple exponential potential as some reference system and treat the short-range perturbation by some type of generalized cluster expansion, similar to that devised in~\cite{GoKr89,Holovko}. 
The mentioned work led, in particular, to a summary of the existing theoretical knowledge on the activity coefficients of electrolytes, extended also the theory 
to mean-spherical approximations and led to applications to transport properties. Special interest was devoted
to the consequences of charge asymmetry and of higher charging for the individual ionic properties. This way new information about the hydration sphere which strongly depends on the charges was obtained
\cite{Krienke2013,EbFeKr21, EbFeKr21_b,EbKrCMP23,Holovko}.
Some results of this collaboration, as well as recent developments, will be discussed here.
As a basis of the present work, we take recent summaries of the state of art about 
systems with exponential interactions and study, in particular, the contributions of the orders $e^4$ and $e^6$ which are most relevant for the systems of our interest~\cite{Yukhn25,EbKrCMP23,Holovko,EbCMP25}. Further we discuss the serious problem, that the Coulomb energy of systems of $N_i$ identical classical charged hard spheres with arbitrary size should not violate Onsager's lower bound 
\begin{equation}
U^C > -\sum_i  N_i \frac{e_i^2}{4 \piup \epsilon k_{\rm B} T d_i} , \quad \epsilon = \epsilon_0 \epsilon_r,
\end{equation}
where $\epsilon_r (T, p)$ is the relative dielectric constant of the solution, and $d_i$ is half the distance between the ions. Further $d_i$ is half the distance between the ions. Onsager found this lower bound by ``Gedanken'' experiments about the total electrical energy of charges in a solution and studies of the known real systems of charges~\cite{Yukhn25,Onsager,Onsager-b,EbCMP25}. Salpeter developed the related model of ion spheres and derived Onsager's results based on classical electrostatics with a prefactor $9/10$ \cite{Salpeter,Salpeter_a}. This seems to be indeed a strict  bound for the electrostatic energy from below, since so far nobody was able to find a counter example. We will check here the relation between the Yukhnovskii--Kelbg theory and Onsager's bound
and show that the bound is observed. We remember that an application of the Yukhnovskii--Kelbg
methods to charged   particle systems in the regions where bound states are formed were given in a foregoing work \cite{EbCMP25}. Here, we study the region of strong electrolytes and fully ionized plasmas
and concentrate on the contributions of order $e^6$ which are, in particular, relevant for systems with unsymmetrical ions as, e.g., the electrolytes CaCl$_2$, LaCL$_3$ and helium plasmas.

\section{Classical systems with exponential interactions}
\label{sec-2}
\subsection{Basic assumptions and model}
\label{sec-2.1}
The theory which Ihor R. Yukhnovskii developed with his adviser Glauberman~\cite{Yukhn02,GlaYuk52} was later extended to the range of higher concentrations in many subsequent works, in particular, by using collective coordinates  \cite{Yukhn02,Yukhn80}. Yukhnovskii and Glauberman used for the description of the forces between charged particles the effective potential, proposed already by Kramers in early quantum theory and Hellman in quantum chemistry. We refer to this class of potentials as exponential potentials and consider, in particular, the simplest form 
equation~\eqref{eq-1}. The prefactor of the potential  in combination with the thermal energy defines a length $\ell$, called thermal Coulomb length, being of the order of a few angstrom.
The exponential potential has a finite height at $r = 0$. 
Glauberman and Yukhnovskii \cite{GlaYuk52,Yukhn80} used this potential  for 
solving equilibrium problems of electrolyte theory, and Kelbg 
developed successful applications to transport problems, in particular, to conductance and viscosity of electrolytes \cite{Falkenhagen,Falkenhagen_b,Kelbg59,Kelbg59_b}. Kelbg extended the potential to the quantum-statistical case and studied important applications to plasmas \cite{Falkenhagen,Falkenhagen_b,Friedman,Barthel}.
In previous work, we used approximations of the nonlinear Debye--H\"uckel and  Onsager--Fuoss type
for calculations of pressure, osmotic coefficient, electrical and free energy \cite{EbFeKr21, EbFeKr21_b,EbKrCMP23}. 
The potentials of average force between the ions $i$ and $j$, $\psi_{ij}$ were separated into a long-range Coulomb potential 
and a short-range potential $V_{ij}'(r)$. Here, we proceed by including an additional exponential part
to the short-range component 
\begin{align}
&\psi_{ij} (r) = V_{ij} + V_{ij}^0, \qquad V_{ij} = \frac{e_i e_j}{4 \piup \epsilon r} [1 - \exp(-\alpha r)],\nonumber\\
&V_{ij}^0 =V_{ij}'(r) + \frac{e_i e_j}{4 \piup\epsilon r} \big[ \exp(-\alpha_{ij} r) - \exp(-\alpha r) \big], \\
&V'_{ij} (r) = \infty \quad {\text{if}} \quad r < R_{ij}, \quad  {\text{else}} \quad V_{ij}'(r)  = 0 . 
\end{align}
This means that we use for generality  species-dependent $\alpha_{ij}$-parameters, but shift for simplicity the corresponding contributions to
the short-range interaction. This is useful, since this way we are avoiding a matrix  theory of screening. To improve the description of the physical situation around an ion, in some older work square well and step
potentials were introduced \cite{Yukhn80,Falkenhagen,Falkenhagen_b,Barthel,Holovko}. The hard-core distances may be  adapted or fixed  as the sum of Pauling radii  \cite{EbFeKr21, EbFeKr21_b,EbKrCMP23}.
Replacing the pure Coulomb potential by an exponential potential permits to take into account several effects, as 
deviations from Coulomb's law due to hydration in solutions and further effects due to Heisenberg uncertainty
in quantum plasmas. In comparison with the Debye--H\"uckel model, the exponential potentials have several important advantages:
\begin{enumerate}
\renewcommand{\labelenumi}{(\roman{enumi})}
	\item The exponential potentials are rather smooth and possess a simple Fourier transform. They are well-suited for the treatment of all screening and collective effects.
	\item  For systems with exponential potential, strict Bogoliubov type expansions exist regarding the plasma parameter \cite{Holovko}.
	\item  The existence of a simple screening equation permits cluster expansions around the 	first exponential type approximation similar to those for Debye--H\"uckel type systems \cite{EbKrCMP23}.
	\item As shown by Kelbg, the exponential potential is also quite useful in applications to quantum plasmas	\cite{Kelbg62,Kelbg63,Kelbg63-b,KelbgHoff64}.
\end{enumerate}

In what follows, we prefer to use exponential potentials with just one $\alpha$. We want to show that this basic case is already very flexible 
and permits, if adapted in the best possible way to the system, to reach excellent descriptions of real systems, already in the first approximation; the needed corrections for including the species-dependence are shifted to the cluster integrals.
The most important advantage of the exponential potential as a reference system is based on the fact that the
Fourier transform is simple:
\be
{\tilde V}_{ij} (k)=  \frac{e_i e_j}{\epsilon} \bigg(\frac{1}{k^2} - \frac{1}{k^2 + \alpha^2} \bigg)
= \frac{e_i e_j}{\epsilon} \frac{\alpha^2}{k^2 \alpha^2 + k^4} .
\ee
We note that our potential includes a weak short-range repulsion having a  Fourier transform and decays at large $k$-values like $k^{-4}$, i.e., much faster than the Coulomb potential. We consider here two potential models as equvalent if the first deviations from
Debye's limiting law are equal. As the first approximation we use the Bogoliubov approach to the pair
correlation function \cite{Yukhn54,Yukhn58,Yukhn80,Yukhn25}
$F_{ij} (1,2) = 1 + g_{ij} (1,2)$ with 
\bea
g_{ij}(1,2) + \beta V_{ij}(1,2) + \beta \sum_k n_k \int V_{ik}(1,3) g_{jk} (2,3) \,\rd {\bf 3} = 0.
\eea
This approximation may be considered as the first term of an  expansion with respect to the plasma parameter \cite{Falkenhagen,Falkenhagen_b,Holovko}.
In some sense, the Bogoliubov equation for the correlaton function $g_{ij}$ is related to the Ornstein--Zernike equation which is the basis of HNC and MSA-related theories~\cite{Barthel,Holovko,Martynov,Evans,Carvalho}.
We have to note, however, that the Bogoliubov type approach is historically 
the first systematic approach to Coulomb systems and is related to more advanced approximations 
\cite{Yukhn25,Barthel,Holovko,Martynov}.

In the general case when $\alpha$ is a matrix, the correlation function is obtained  by solving matrix-integral equations \cite{Falkenhagen,Falkenhagen_b,EbFoFi17,EbFoFi17-b}. In the simplest case, provided all $\alpha$ are equal, or we use this case as a reference system, 
the solution is particularly easy and reads
\bea
g_{ij} = - \frac{\beta e_i e_j \alpha^2}{4 \piup \epsilon r \big(p^2 - s^2\big)} \big[\exp(-pr) - \exp(-sr)   \big] .
\eea
The parameters s and p are solutions of a 4th order polynomial. At very small densities, the solutions are
$s = \alpha$ and $p = \kappa = 1/r_D$.
At intermediate densities, we have \cite{Falkenhagen,Falkenhagen_b,Holovko}
\bea
 \frac{p}{\alpha} = \frac{1}{2} \left(\sqrt{1 + 2 \frac{\kappa}{\alpha}} -\sqrt{1- 2 \frac{\kappa}{\alpha}}\right), 
 \qquad\frac{s}{\alpha} = \frac{1}{2} \left(\sqrt{1 + 2 \frac{\kappa}{\alpha}} +\sqrt{1 - 2 \frac{\kappa}{\alpha}}\right).
 \label{eq-8}
\eea
The Coulomb energy is determined by the charge density around an ion and we find
the result  \cite{Holovko} 
\bea  
U_{\rm C} =  \sum_i N_i u_{i,{\rm C}} = - \frac{V}{8 \piup }  \frac{\kappa^3}{\sqrt{1 + 2 \kappa/\alpha}} 
= - \frac{V}{8 \piup \epsilon} \frac{\kappa^3}{(1 + 2 \gamma/\alpha)}.
\eea
Here, $\gamma$ is a variable defined by $\kappa = 2 \gamma(1 + \gamma/\alpha)$ which provides a form of the equations, which is convenient for comparisons with the DH and the MSA theory \cite{Holovko}.
For  large densities with $2 \kappa/\alpha >1$, i.e., $2 /\alpha  > r_D$, we find a different solution \cite{Holovko}
\begin{align}
&g_{ij}(r; \kappa>\alpha/2) = - \frac{\beta e_i e_j}{8 \piup \epsilon r} \frac{\alpha^2}{q t} \exp(-qr) \sin(t r) ,\\ 
&t/\alpha = (1/2) \sqrt{2 (\kappa/\alpha) -1}, \quad q/\alpha = (1/2) \sqrt{1 + 2 \kappa/\alpha}.
\end{align}
Note that the transition point makes no problems since for $2 \kappa/\alpha = 1$, the limit is a well defined integrable function.  Note that this transiton was also studied  in a recent monograph \cite{AlMaMono}. We note further that the result for the electrical energy for $2 \kappa/\alpha >1$ is the same as for lower  values of $\kappa/\alpha$, i.e., in this order there is no qualitative change at the point of transition. 

For the free energy,  we get the pressure and the activity coefficient  by using a charging process according to Yukhnovskii and Kelbg 
for the exponential potential 
\cite{Yukhn02,Kelbg59,Kelbg59_b}: 
\begin{align}
\beta F_{\rm KY} &= \beta F_{\rm id} - \frac{\kappa^3}{12 \piup} \tau\left(\frac{\kappa}{\alpha}\right) , \quad \beta p = n - \frac{\kappa^3}{24 \piup} 
\varphi\left(\frac{\kappa}{\alpha}\right), 
\label{eq-12}\\
  \varphi(x) &= \frac{1}{x^3} \left(\frac{1}{4} X^3 - \frac{2}{3} X - \frac{3}{4 X}\right), \quad 
\tau(x) = \frac{3 \alpha^3}{4 \piup \kappa^3}\left(Y^2 - 2 Y + \frac{2}{3} Y^3 - \frac{1}{2} Y^4 + \frac{5}{6} \right), \nonumber \\
\ln f_i &= - \frac{e_i^2}{8 \piup \epsilon k_{\rm B} T} \left(\frac{Y - 3}{X} \right), \quad  Y = 2 + X, \quad  X = \sqrt{1 + 2 \frac{\kappa}{\alpha}}.  \qquad
\label{eq-13}
\end{align}
Note that in the oscillation region  we have $X > \sqrt{2}$ and so far it remains unclear what this means regarding the thermodynamics.

\begin{figure}[htbp]
\begin{center}
\includegraphics[scale=0.3]{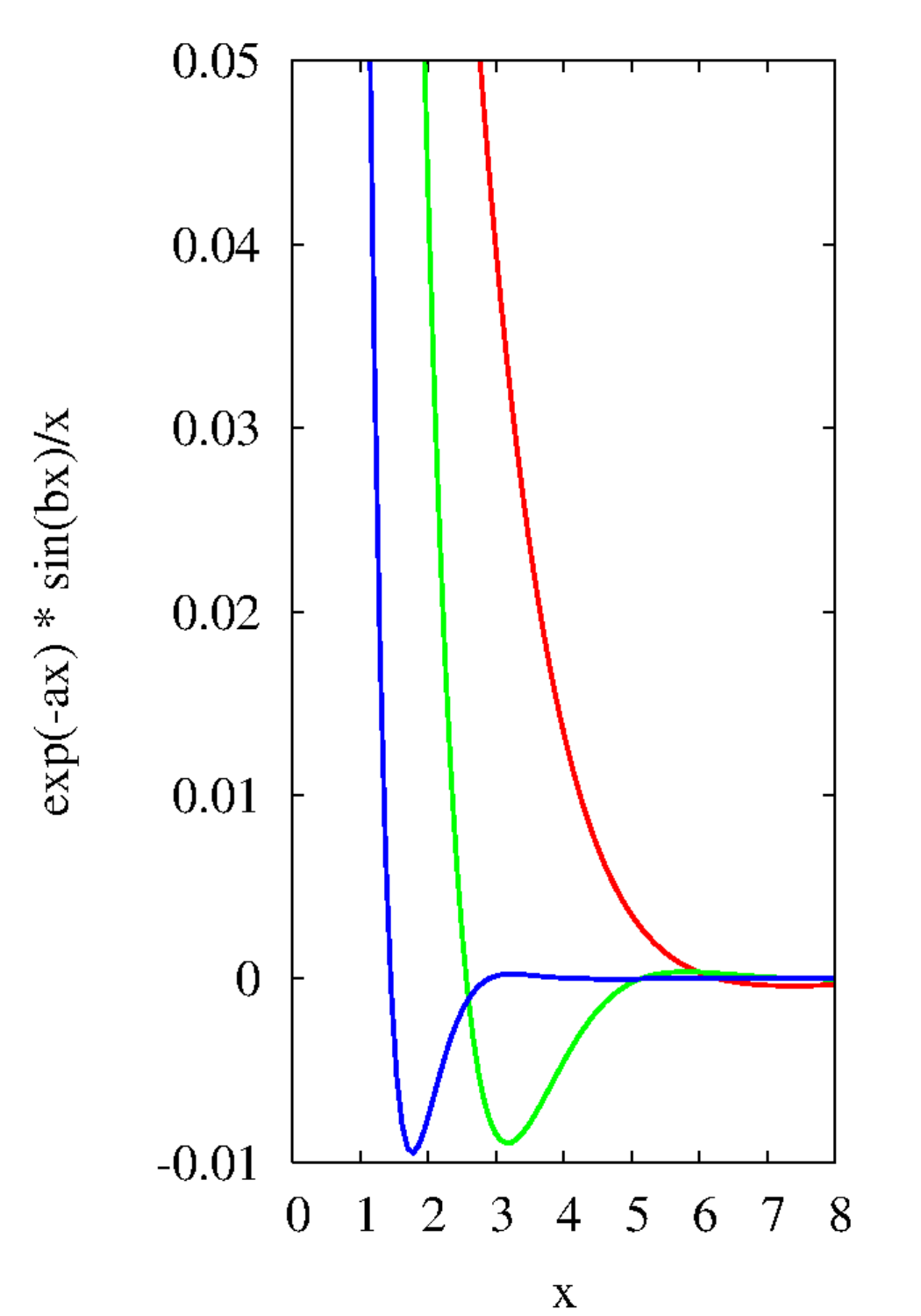}
\caption{(Colour online) The function characterizing the decay of the opposite charge density around an ion 
in the Yukhnovskii--Kelbg  theory
for the density parameters $\kappa / \alpha = 1$ (exponentially decaying red curve above), for $\kappa / \alpha=2.5$ (green curve with minimum below on r.h.s.) and for $\kappa / \alpha = 9$ (blue curve with minimum below on l.h.s.). We note that the negative parts correspond to the second charge shells with the same sign as the central ion, which do not appear in the ion sphere theory.}
\label{fig-2}
\end{center}
\end{figure}

\begin{figure}[htbp]
\begin{center}
\includegraphics[scale=0.3]{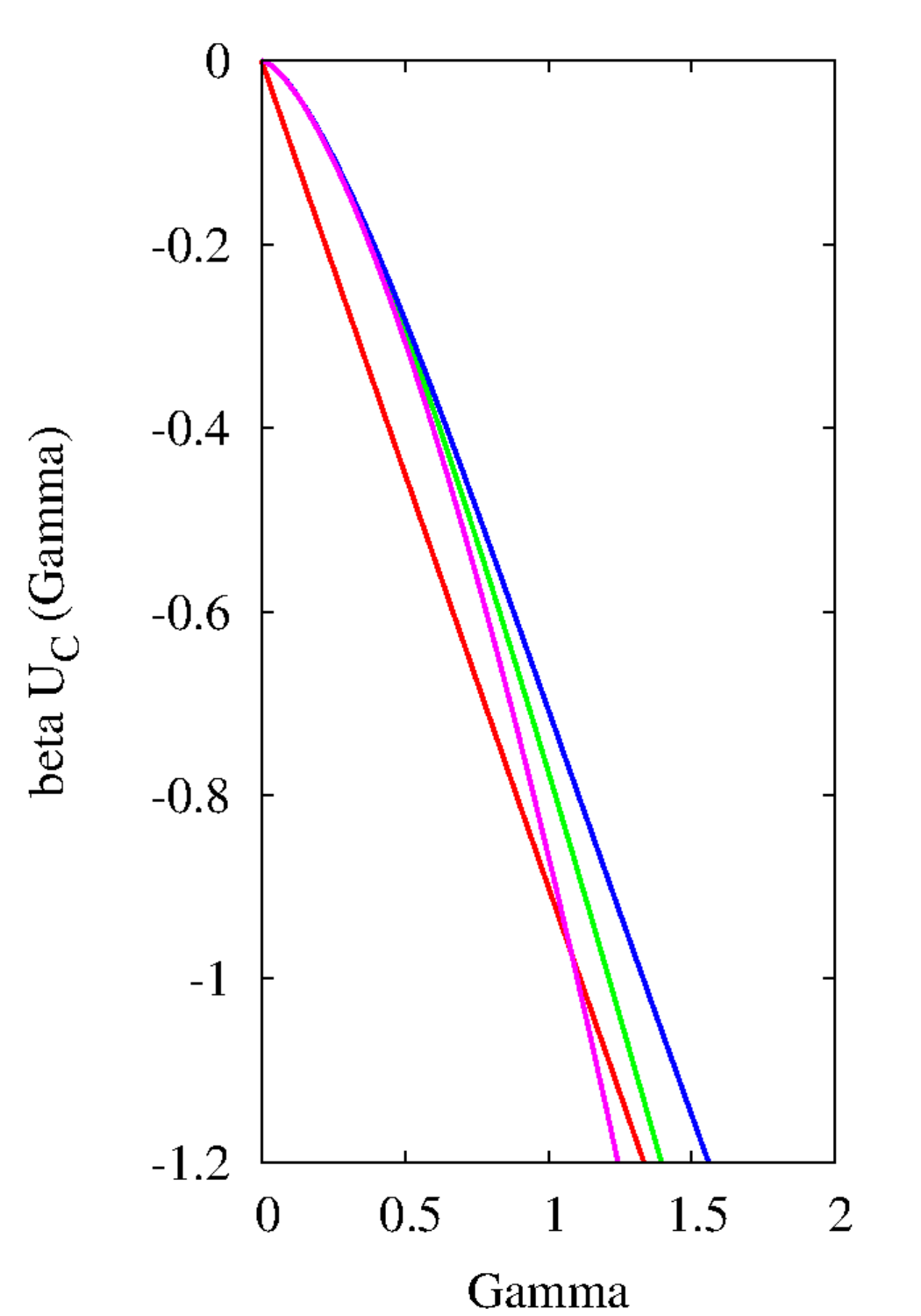}
\caption{(Colour online) The cubic root law of Onsager (rectilinear red curve below) in comparison with Debye's quadratic root law 
(turquoise curve crossing first Onsagerws law) and the results of Yukhnovskii and Kelbg (blue curve above) for the Coulomb energy for NaCl.
The Debye--H\"uckel theory (in green) crosses also the Onsager bound at higher densities. Note that the 
Yukknovskii--Kelbg theory (blue curve above) observes the bound due to a more moderate increase like $n_i^{1/4} \sim \Gamma^{3/4}$. 
}
\label{cubicroot}
\end{center}
\end{figure}

We  may compare all these results with the known formulae for the Debye--H\"uckel model~\cite{Falkenhagen,Falkenhagen_b,Harned,Barthel}.
For such a comparison of the Debye--H\"uckel reference system with the exponential 
potential model, we may define an $a$ parameter conjugated to our $\alpha$  parameter by $a \sim \alpha^{-1}$.  Corrections by additional short-range and a hard core part of the potential may be included by the second virial contribution~\cite{Barthel} as will be discussed below. The case of large $\kappa/\alpha > 1/2$ which corresponds to oscillating potentials~\cite{Holovko} also remains to be discussed. 
In spite of the fact that the formulae obtained with the exponential potential has been very successful in describing data, there still remain  open problems. One of them is the relation to the Onsager--Salpeter bound including ion sphere effects which may play a role at larger densities. The graphical representation of 
the Coulomb energy formulae of Yukhnovskii and Kelbg, which we demonstrated in figure~\ref{fig-2},
shows that the low density behaviour of the Coulomb energy is detetermined by 
the screening parameter $\kappa \sim n_i^{1/2}$ and the screening length $r_D \sim n_i^{-1/2}$. The Yukhnovski--Kelbg screening means in terms of physics that at higher densities there are  two screening lengths $1/s$ and $1/p$, one being a bit smaller and one being a bit larger than $r_D$. This is already a first step
to screening at higher densities. At rather high densities, these lengths compete with the length $d_i$, which is half the average distance between the ions. In order to explain the situation in terms of the spirit of Onsager and Salpeter, let us think about multicharged ions as Ca$^{2+}$, Fe$^{3+}$  in an ionic solution, or He$^{2+}$ and Li$^{3+}$ in a solar plasma. According to Onsager and Salpeter, the most relevant quantity
 at higher density is the charge density. In Salpeter's electrostatic model, he assumes that the charge 
around an ion forms a sphere filled with opposite charge \cite{Onsager,Onsager-b,Salpeter,Salpeter_a,Hauge}. The radius of this ``ionic sphere'' is $d_i$, i.e., half of the ionic distance, since beyond it the next sphere already begins. 
According to Salpeter, the high-density behaviour is determined by this effect with 
leading terms of the order $n_i^{1/3} \sim d_i^{-1} $. In the model of Onsager and Salpeter, the charge density tends to zero if $r$ approaches $d_i$ and is near  zero beyond. 
The result of Onsager and Salpeter for the Coulomb energy betrween the central ion and the cloud is in accordance with the laws of electrostatics~\cite{Onsager,Onsager-b,Salpeter,Salpeter_a}. According to Salpeter's model, the charge density inside
the sphere is uniform. The potential as well as the Coulomb energy of the ion are as follows: 
\begin{align}
\Phi_i(r) &= \frac{e_i}{4 \piup \epsilon r} \left( 1 - \frac{3 r}{2 d_i} + \frac{r^3}{2 d_i^3} \right), \quad 
\Phi_i(r) = 0 \quad {\text{if}} \quad r > d_i,\\ 
u_{i,{\rm C}} &= - \frac{9}{10} \frac{e_i^2}{4 \piup \epsilon d_i}. \qquad 
\end{align}
According to Onsager, this is a strict lower bound for the Coulomb energy per ion \cite{Onsager,Onsager-b}.
He has given a proof, that there is no configuration for ions having a hard core of arbitrary size and shape, which  a lower Coulomb energy has got. The shape of the charge density according to the 
Yukhnovskii--Kelbg theory is shown in figure~\ref{fig-2} and the corresponding Coulomb energy in figure~\ref{cubicroot} in comparison to Debye's law.

Asuming that the Onsager--Salpeter theory describes the high density limit in a correct way, we 
see that the Debye--H\"uckel and the Yukhnovskii theory might have a  problem with the 
large distance decay of the charge density at higher ion concentrations. At  $\kappa > \alpha/2$, the charge density is in the YK theory given by
\bea
\sigma_i^{\rm YK} = - e_i  \frac{2 \kappa^2 \exp(-qr) }{ \sqrt{4 (\kappa/\alpha)^2 -1}} \,\frac{\sin(tr)}{r}.
\eea

In other words, theories predicting lower Coulomb energies have a problem with Onsager's bound. 
We have drawn in figure~\ref{cubicroot} the Coulomb energy in comparision to the Debye--H\"uckel and show that the DH theory tends to violate the Onsager bound, while the 
YK theory does not. Due to the more moderate increase as $n_i^{1/4}$ with the density,
the YK curves never cross the bound of Onsager--Salpeter. Note that we concentrate here on the effect of
Coulomb forces and do not consider explicitely the short-range contributions. 
We will discuss in a forthcoming work how the YK  theory and the Onsager--Salpeter theory may be 
combined and the way how caging effects may be included. 

The transitions, which we discussed here in thermodynamic functions are rather smooth as seen in  figure~\ref{cubicroot}.
On the other hand, we have seen that the structural effects, observed, e.g., in the correlation functions, show a qualitative transitions
to an oscillating components. However, the question about observations of these structural effects remains open.  On the other hand, the $n^{1/3}$ effects or in other words a 
linear dependence on the plasma parameter $\Gamma = \ell/d_i \sim n_i^{1/3}$
($d_i$ is a mean distance between ions) was already sometimes predicted
in the studies of electrolytes and plasmas \cite{Harned,Stolzmann,Onsager,Onsager-b}. 
 
In order to summarize we conclude the following. 
The traditional Debye--H\"uckel screening is supplemented at higher concentrations 
by an additonal strong screening effect which has an effect on the thermodynamic functions. 
This is modelled as Onsager--Salpeter caging of the charges by forming an ion sphere of size~$d_i$. Here, $d_i$ is half the average distance to the next ion of the same charge. The ``caging'' of the ion together with its inner neutralizing cloud means that outside the sphere $r = d_i$, defined by $V = 4 \piup N_i d_i^3/3$,  the Debye--H\"uckel ``screening cloud'' is practically destroyed and replaced by an ``ion sphere'' screening.  The effect is in both cases similar, the central charge is neutralized by a cloud with opposite charge. The structure is, however, completely different. The Debye screening cloud is large and exponentially deacaying, the Onsager ion sphere is small with a nearly constant charge density filling a sphere with the radius $r =d_i$. Note that the candidates for a closer experimental study of the effects that we discussed here are, e.g., Ca$^{2+}$ and 
La$^{3+}$ ions in solutions or  Li$^{3+}$ and
C$^{6+}$ ions in plasmas, which show much stronger Coulomb effects than single-charged ions and electrons. 
Friedman has checked the influence of the term~$e^6$ by a careful comparison of the the cluster theory with experiments 
for LaCl$_3$ electrolytes at very small concentrations. In our view, this and a careful check of the cubic root $n_i^{1/3}$ laws by real experiments on electrolytes and plasmas belongs to the most interesting open questions. The numerical experiments on high density classical Coulomb systems seem to provide evidence that at high density the $n_i^{1/3}$
laws appear in thermodynamics \cite{DeWitt,KKER86,EbFoFi17,EbFoFi17-b}, a careful numerical check of the effects of charge asymmetry is not known to us.

\subsection{Higher approximations and cluster expansions}
\label{sec-2.2}
Generalizations of Bogoliubov's correlation function and the corresponding thermodynamic functions were given in \cite{Yukhn54,Falkenhagen,Falkenhagen_b,Tjablikov}. The second approximation of the Bogoliubov expansion 
for the pair correlation function $F_{ij} (r)$ reads
\cite{Yukhn58,Falkenhagen,Falkenhagen_b}:
\bea
F_{ij} (1,2) = \exp[g_{ij} (1,2)]\bigg[1  + \frac{1}{2!} \int \rd {\bf 3} \big(g_{ik}^2 g_{jk} + g_{ik} g_{jk}^2 \big) 
 + \frac{1}{2!} \int  \rd {\bf 3} \,\rd {\bf 4 } \big( g_{jk} g_{kl}^2 g_{lj} \big) + \ldots\bigg].
\eea
The correlations have, for unsymmetrical systems, at large distances and $r > 2/\alpha > 1/\kappa$ a tail which behaves like $\exp(-\kappa r)$ \cite{Falkenhagen,Falkenhagen_b} 
\bea
F_{ij} \simeq \frac{2 \piup^2}{\kappa^2} \ln 3 \sum_{k,l} n_k n_l \ell_{ik} \ell_{jl} \ell_{kl}^2 \frac{\alpha^2}{(p^2 - s^2)} \exp(-\kappa r) 
+ O\bigg(\frac{\exp(-\kappa r)}{r}\bigg).
\label{eq-18}
\eea  
The hehaviour of the new long-range tail may lead to strong correlations, e.g., in the case that in a plasma two protons are 
imbedded between He$^{2+}$ or Li$^{3+}$ ions. This may be relevant, e.g., for problems of fusion. 
This tail is still to be explored for the region of oscillations. We note that the tail is also responsible for the replacement of the factor $\ln 2$ by a factor $\ln 3$ in the pressure as shown later in equations~\eqref{eq-19},~\eqref{eq-28},~\eqref{eq-29}.

An exact result for the osmotic pressure including the long-range term in the pair distribution reads~\cite{Falkenhagen,Falkenhagen_b}
\bea
\frac{p}{p_{\rm id}} = 1 - \frac{2 \piup}{3 n} \sum_{ij} n_i n_j \ell_{ij} \bigg(1 - \frac{2 \kappa}{\alpha} + \ldots \bigg) 
- \frac{\piup}{3 n} \sum_{ij} n_i n_j \ell_{ij}^3 \bigg(\ln \frac{\kappa}{\alpha} + \ln 3 + 2C - \frac{11}{6} \bigg).
\label{eq-19}
 \eea
For a more detailed study of the third order $O(e^6)$ of the thermodynamic functions and, in particular, the Coulomb energy, we study now the integrals  
\bea
u^{\rm C}_3(\kappa)= -\frac{k_{\rm B} T}{2} \sum_{ij} n_i n_j  \int \rd {\bf 2} \frac{\ell_{ij}}{r} g_{ij}^2(r),\quad \int_0^{\infty} \rd x \frac{\exp(- ax) - \exp(-bx)}{x} = \ln  \frac{b}{a} .
\label{eq-20}
\eea
This way for the order $e^6$ in the density of Coulomb energy of the ions,  we get
\bea
u^{\rm C}_3(\kappa)= - k_{\rm B} T \frac{\piup}{8} \sum_{ij} n_i n_j \ell_{ij}^3 \big[\ln (2p/\alpha) + \ln (2s/\alpha)- 2 \ln (s + p)/\alpha\big].
\eea
Introducing the indices $[ij]$ for a species-dependence of the $\alpha$ parameter 
in the lowest order, for the third order 
 of the Coulomb energy of the ions,  we get
\be
u^{\rm C}_3 (\kappa)= - k_{\rm B} T \frac{\piup}{2} \sum_{ij} n_i n_j \ell_{ij}^3 \ln \frac{(2 \kappa/\alpha_{ij})} {\sqrt{1 + 2 \kappa/\alpha_{ij}}}.
\ee
The Coulomb energy provides in the third order relevant contributions to the thermodynamic functions.
For the third order contribution to the osmotic pressures, using the virial formula,  we get, e.g.:
\bea
p_3 (\kappa)= \frac{u^{\rm C}_3 (\kappa)}{3} = - k_{\rm B} T \frac{\piup}{6} \sum_{ij} n_i n_j \ell_{ij}^3 \ln \frac{(2 \kappa/\alpha_{ij})} {\sqrt{1 + 2 \kappa/\alpha_{ij}}}.
\label{eq-23}
\eea
The logarithmic term $\ln(\kappa/\alpha_{ij})$ is in agreement with the exact result equation~\eqref{eq-19}, the following constant is only an approximation.
 For larger $\kappa$ values $2 \kappa > \alpha$, we have to recalculate the pressure using the periodic $g$  function according to equations~\eqref{eq-12},~\eqref{eq-13} and get
\bea
p_3 (\kappa)= - k_{\rm B} T \frac{\piup}{6} \sum_{ij} n_i n_j \ell_{ij}^3 \frac{1}{(2 \kappa/\alpha_{ij} -1)(2 \kappa/\alpha_{ij} +1)} 
\ln \bigg[1 + \frac{(2 \kappa/\alpha_{ij} -1)}{2 \kappa/\alpha_{ij}+1}\bigg].
\label{eq-24}
\eea
At the point of transition, this gives
\bea
p_3 (\kappa = \alpha/2)= - k_{\rm B} T \frac{\piup}{24} \sum_{ij} n_i n_j \ell_{ij}^3.
\label{eq-25}
\eea
Note that this value is different from the limit performed from below which gives
the prefactor $\piup \ln(2)/12$ instead of $\piup/24$  and this means that we obtain a small gap at the transition point. It remains unclear so far, what is the physical 
 interpretation of this gap, due to a discontinuity in the 2nd order. Anyhow, a phase transition or any anomaly at this $\kappa$-value has never been observed so far in experiments or numerical simulations. 
We should not overestimate our result about a discontinuity of the constant, since the constant following from equation~\eqref{eq-24} is not exact, as to be seen by comparison with equation~\eqref{eq-20}. As known from earlier work, the  integral terms in equation~\eqref{eq-19} which we neglected, contribute to the constant~\cite{Falkenhagen,Falkenhagen_b,KKER86}. Therefore, as far as we are not able to treat all terms contributing to the constant in the logarithmic term and include the integrals in equation~\eqref{eq-19} exactly, we should consider the question about the gap in the logarithmic order as open. 
In conclusion, our strategy is as follows. We adapt the free constant in equations~\eqref{eq-23},~\eqref{eq-24} to the exact result equation~\eqref{eq-20} and get
\bea
p_3 (\kappa)= \frac{u^{\rm C}_3 (\kappa)}{3} = - k_{\rm B} T \frac{\piup}{3} \sum_{ij} n_i n_j \ell_{ij}^3 \ln \frac{(\gamma_2 \kappa/\alpha_{ij})} {\sqrt{1 + 2 \kappa/\alpha_{ij}}},\label{eq-26}\\
\gamma_2 = \ln 3 + 2C = 4/3 \simeq 2.4960, \qquad {\text{if}} \quad \kappa < \alpha/2.
\label{eq-27}
\eea
Further we neglect the small gap we obtained between equations~\eqref{eq-24},~\eqref{eq-25} and find adapting the uncertain constant to the exact value $\gamma_2$, for the third order contribution to the osmotic pressures 
\bea
p_3 (\kappa)= - k_{\rm B} T \frac{\piup}{6} \sum_{ij} n_i n_j \ell_{ij}^3 \frac{\sqrt{2} \gamma_2}{4 \kappa^2/\alpha_{ij}^2 -1}  \ln \frac{(4 \kappa/\alpha_{ij})}{1 + 2 \kappa/\alpha_{ij}}  \quad {\rm if}\quad \kappa > \alpha/2.
\label{eq-28}
\eea
 Note, that this is a different analytical function which generalizes the function defined by 
 equations~\eqref{eq-27},~\eqref{eq-28} which continue the exact expression~\eqref{eq-20}.
 The logarithmic contribution $O(e^6)$ is relevant for the thermodynamics of strong electrolytes, in particular, in the case of asymmetries like for solutions of 
CaCl$_2$, K$_2$SO$_4$, LaCl$_3$, \ldots\,.
The Yukhnovskii--Kelbg approximation provides a most important part of the osmotic pressure, although we should also include the known contributions of the short-range interactions like hard cores of the ions \cite{Falkenhagen,Falkenhagen_b,Holovko}. For example, we may use the model of charged spheres with adaptive
non-additive contact distances $R_{ij}$ \cite{EbFeKr21, EbFeKr21_b,EbKrCMP23} or  a corresponding exponential potential.
Solving the YK problem for this exponential potential, we arrive at the electrical energy density $u^{\rm C} (\kappa)$ and find the osmotic pressure
\bea
\beta p = \sum_i n_i - \frac{u^{\rm C}(\kappa)}{3} + \frac{\piup}{3} \int_0^{\infty} \rd r \,r^2 
\frac{\partial V_i'(r)}{\partial r} F_{ij} (r),
\label{eq-29}
\eea
which leads to 
\bea
\beta p =  \sum_{i} \bigg\{n_i - \frac{u_i^{\rm C}(\kappa)}{3}  - \frac{\piup}{3} \sum_j \bigg[\exp\bigg( - \frac{\beta e_i e_j \alpha^2}{4 \piup \epsilon R_{ij} (p^2 - s^2)}\bigg) \big(\re^{-pR_{ij}} 
- \re^{-s R_{ij}}  \big) -1\bigg] \bigg\}. \nonumber
\eea
This is the most easy way to include the hard-core contributions.
The contact distances $R_{ij}$ are considered as adaptable parameters. 
We note the close relation between the osmotic coefficient in the YK theory and the MSA in the form proposed by Henderson, Smith and others \cite{EbFeKr21, EbFeKr21_b,EbKrCMP23}.

In a more advanced procedure, the long-range part $V_{ij}$ is identified with the 
averaged exponential potential but the species-dependent part with a parameter $\alpha_{ij}$, which is relevant in the case of quantum plasmas, is includecd into the short-range parts by the splitting
\begin{align}
 \psi_{ab} &= \frac{e_i e_j}{ 4 \piup \epsilon r}  [1 - \exp(-\alpha_{ij} r)] +V_{ij}'(r)
 =  \frac{e_i e_j}{4 \i \epsilon  r}  [1 - \exp(-\alpha r)]  \nonumber \\
  &+ \left\{\frac{e_i e_j}{4 \piup \epsilon r} [\exp(-\alpha r) - \exp(-\alpha_{ij} r)] +  V_{ij}' (r)\right\},
 \end{align}
 where  for technical reasons we consider the second contribution as the short-range interaction
 and the first contribution as the long-range part.
Within this model, we may apply the standard cluster expansion technique \cite{Yukhn80,Falkenhagen,Falkenhagen_b,Friedman,Holovko} which is based on the classical work of Mayer, extended by Friedman~\cite{Friedman}. In most detail this technique was developed with applications to electrolytes in  the collaboration Yukhnovskii--Kelbg~\cite{Falkenhagen,Falkenhagen_b,Yukhn80,YukhnC26,Holovko}. 
For small Bogoliubov parameters, for higher densities in the 2nd approximation  we find the cluster expansion \cite{Yukhn25,Falkenhagen,Falkenhagen_b,Holovko}
\begin{align}
F_{ij}(1,2) &= \exp[- \beta V_{ij}^0 + g_{ij} (r)] \bigg\{1 +\sum_k n_k \int \rd {\bf r}_3 \big[ \Phi_{ik} \Phi_{jk}
 + \Phi_{ik} g_{jk} + g_{ik} \Phi_{jk}\label{eq-31}\nonumber\\ 
  &+ \exp(g_{ik} + g_{jk} - \beta V_{ik}^0 - \beta V_{jk}^0 \big]\bigg\},\\ 
 \Phi_{ik} &= \exp(g_{ik} - \beta V_{ik}^0) -1 - g_{ik}.\nonumber
\end{align}
General expressions for the free energy and other thermodynamic functions follow from statistical thermodynamics for the cluster contributions by a charging procedure. 
For the model of hard charged spheress with equal diameters $\sigma = R_{ij}$, the so-called RPM-model, follows the well-known result
\bea
F  = F_{\rm id} - k_{\rm B} T V \bigg\{\frac{\kappa^3}{12 \piup} \tau(\kappa \sigma)  + \frac{1}{2} \sum_{ij}  n_i n_j \int 
\rd {\bf r}_1 \rd {\bf r}_2 \left[\Phi_{ij}(1,2) - \frac{1}{2} g_{ij}^2\right] + \dots \bigg\},
\eea
where the Debye--H\"uckel function is well-known \cite{Barthel,EbKrCMP23,EbCMP25}.
For higher concentrations, one may add further terms of the cluster expansion \cite{Yukhn80,Falkenhagen,Falkenhagen_b,Friedman,Barthel,EbKrCMP23}.
Summarizing our state for electrolytic systems we conlude that in the theory of concentrated electrolytes,  several problems are still open, similar to the ones  in the theory of dense nonideal plasmas \cite{EbFoFi17,EbFoFi17-b}.
Closing this chapter on concentrated strong electrolytes in aqueous solutions, the pioneers Harned and Owen 
write, describing the statements of Onsager and Fuoss: ``At high concentrations, the distribution function must be a damped periodic function of the interionic distances'' \cite{Harned}. Our own findings agree with this statement (see the previous sections and \cite{Holovko}). We refer, in particular, to the result obtained from the correlation function at high densities [see equation~\eqref{eq-8}]. Our results  say that for $\kappa/\alpha > 1/2$, the correlation function of a Coulomb system with exponential interaction shows a damped oscillation (see section~\ref{sec-2.1}). However, no phase transitions are connected with these oscillations,  they are structural transitions. Evidently, the thermodynamic functions are smooth but we cannot exclude that the derivatives show peculiarities. The open questions are connected with the behaviour of the integral contributions to the correlations in equation~\eqref{eq-18}.
Our statements are closely related to earlier findings of Martynov, Evans, Carvalho and others about the decay of correlations in the restricted primitive model of ionic fluids \cite{Martynov,Evans,Carvalho}.
These auhors showed that the longest range decay of each of the total correlation functions
is determined by a single exponential decay length. 
Our statements are closely related to earlier findings of Carvalho and Evans about the decay of correlations on the model of the restricted primitive model of ionic fluids \cite{Carvalho}. These authors observed already for the RPM model transitions from a monotoneous decay of the charge density to an oscillatorry decay for 
increasing ion density. It was suggested to us to compare our result for the transition with similar results for the restrictive primitive model. In particular, when one puts $\alpha = 2/\sigma$ with $\sigma$ being the diameter of the ion, the transition point $\kappa/\alpha = 1/2$ obtained in our consideration for the 
change from exponential decay of correlations to damped oscillation transforms to $\kappa \sigma = 1$ which is close to the results obtained for the restrictive primitive model by other authors by different theories.
For example, the mean-spherical approximation~\cite{Outwaite} gives for the point of transition 
$(\kappa \sigma)_c = 1.229$, while in the recent simulations there was obtained $(\kappa \sigma)_c = 1.37$~\cite{Cats}.
The close relation between the RPM and the YL model with equal $\alpha$-parameters suggests us to call our model a restricted YK model (RYK). 
In any case, we may guess that an oscillatory decay of correlations is connected with the Coulombic tail rather than is model-specific as already guessed by Harned and Owen~\cite{Harned}.

\section{Cluster expansions for quantum plasmas} 
\label{sec-3}
\subsection{Cluster expansions for exponential potentials} 
\label{sec-3.1}
Kelbg's basic idea was that for the nondegenerate case, the quantum effects in Coulomb systems may be described by effective potentials. 
Introducing such effective potentials, in particular, the Kelbg potential \cite{Kelbg62}, many methods developed for the classical case are still working. In particular, the cluster expansions remain essentially the same as in the classical case \cite{KKER86,EbFoFi17,EbFoFi17-b}. Here, we use the exponential potential as a good approximation for Kelbg's effective quantum potential \cite{EbCMP25}. We note that the circles and lines of graphical representations are slightly different from the classical Mayer expansions. The lines now represent Kelbg interactions instead of Coulomb interactions,
and instead of classical Boltzmann factors appear Slater functions \cite{EbRoMDPI,EbFoFi17,EbFoFi17-b}.
The whole formalism for the quantum case is very close to the Yukhnovskii--Kelbg theory.
In the quantum case, the $\alpha$-parameter of the exponential potential may be introduced in different ways. For example, we may fit the height  at zero distance by $\alpha_{ij} = \sqrt{\piup} / \lambda_{ij}$ or fit the long-range tail of the 
Fourier transform which leads to $\alpha_{ij} = \sqrt{6} /\lambda_{ij}$ \cite{EbFoFi17,EbFoFi17-b}. 
As proposed in~\cite{EKK76,EbCMP25} here  we use a different  definition which reproduces the exact thermodynamic functions up to the first deviations from the Debye contributions to thermodynamic functions. This condition, which is most appropriate for the quantum statistics of plasmas \cite{EKK76}, gives 
$\alpha_{ij} = \sqrt{\piup} / 4 \lambda_{ij}$. In the case that we prefer to work in the screening procedure with just one averaged quantum length, we recommend to define $\alpha$ by   
\bea
\frac{1}{\alpha} = \frac{\sqrt{\piup}}{4} \frac{\sum_i \sum_j e_i^2 e_j^2 \lambda_{ij}}{\sum_i \sum_j e_i^2 e_j^2},
\quad \lambda_{ij} = \frac{{\bar h}}{\sqrt{2 m_{ij} k_{\rm B} T}},
\label{pairav1}
\eea
where $\lambda_{ij}$r
 is the relative thermal wave length of a pair $i,j$.
This provides the correct Debye root law and correct first deviations $O(\kappa \lambda)$ from it.
Due to the similar mathematical structures in the classical and in the quantum case,
 most statistical relations have the same form, but the correlation functions  now depend  on the thermal wave length $\lambda_{ij}$. The simplest way to introduce the screening potentials is to express the direct potential by a screened one by iterating the Bogoliubov equation 
\bea
{\tilde V}_{ij}(1,2)  = V_{ij} (1,2) -  \sum_k n_k \int \rd\mathbf{3}\, V_{ik}(1,3) {\tilde V}_{kj} (2,3),
\eea
where $V_{ij} $ is the quantum exponential potential. The screened potential ${\tilde V}(1,2)$ is connected with the pair distribution and correlation function in lowest order by $g_{ij} (1,2) = -\beta {\tilde V}_{ij} (1,2)$.
For better approximations, we have to include further short-range terms and higher terms in the Bogoliubov parameter. Transforming the classical formula $F_{ij} = \exp\big(g_{ij} - \beta V_{ij}^0\big)$, we see that in the quantum case the classical Boltzmann factor is to be replaced by the 
so-called Slater function which expresses relative probabilities and is defined by the trace over the binary exponential operator \cite{KKER86}
\bea
 \exp\bigg(- \beta V_{ij}'  - \frac{\beta e_i e_j}{4 \piup \epsilon r} \bigg) \longrightarrow S_{ij}(r)  = 2! \Lambda^{6}\, {\rm Tr}\, [\exp(-\beta 
{H}_{ij} )].
\eea
This may be considered as the definition of the effective short-range potential for quantum plasmas and for the pair distribution, expressed in terms of the ``classical'' formula~\eqref{eq-31}. Thus we obtain
the short-range part $V_{ij}'$ by the Slater function using 
\bea
\exp(-\beta V_{ij}') \rightarrow S_{ij} (r) \exp[  \beta e_i e_j / (4 \piup 
 \epsilon r)]
\eea
and get the pair distribution including screening 
\bea
F_{ij} = S_{ij} (r) \exp \big[g_{ij}(r) + \beta e_i e_j / (4 \piup \epsilon r)\big].
\eea
This gives the final form of the first order ditribution  
\bea
F_{ij} = S_{ij} (r) \exp \left[ -\beta \frac{e_i e_j}{4 \piup \epsilon r} (\exp(-p r) - \exp(-\alpha r) -
 1) \right],
\eea
where we have in zeroth approximation according to Kelbg with the quantum statistical $\alpha$:
\bea
S_{ij} (r)  \simeq \exp\left\{ - \beta e_i e_j [1 - \exp(-\alpha r)] / 4 \piup \epsilon r \right\}.
\eea
For higher approximations, we may use equation~\eqref{eq-31} with \cite{KKER86}
\bea
 \phi_{ij} = S_{ij}(r) \exp\{g_{ij} + \beta [e_i e_j (1 - \re^{-\alpha r})r]/ 4 \piup \epsilon r \} - 1 - g_{ij}.
 \eea
The only difference to the classical case is  that we expressed the short-range potential by the pair Slater functions 
using the  relations between quantum and classical statistics \cite{KKER86}. 
This way by using the YK- model we finally get   a quantum generalization
of the cluster expansion of the free energy
\begin{align}
F  &= F_{\rm id} - k_{\rm B} T V \bigg\{\frac{\kappa^3}{12 \piup} \tau\bigg(\frac{\kappa}{\alpha} \bigg)  \nonumber \\ &+ \frac{1}{2}  \sum_{ij} n_i n_j \int \rd {\bf r} \bigg[ S_{ij}(r) \exp\left(g_{ij} + \beta \frac{e_i e_j}{4 \piup \epsilon r}  (1 - \re^{-\alpha r})
\right)
 -1 - g_{ij}  - \frac{1}{2} g_{ij}^2 \bigg] + \cdots \bigg\}.
\end{align}
So far, these relations are already  known \cite{EKK76,KKER86,AlMaMono} so that we have a direct access to the results in the form of an extended limiting law and the cluster integrals as given above \cite{EbRoCPP25}.
The characteristic parameter characterizing the screening effects depends on $\kappa$ or more correctly, 
on the Bogoliubov parameter $\mu =  \ell \kappa \sim O(n_i^{1/2})$.

\subsection{Improved representations of the ring contribution}
\label{sec-3.2}

For Coulomb systems, the divergencies due to long-range interaction may be avoided by summing up diagrams of ring and chain type. At lower densities and higher temperatures, 
the ring contributions provide an essential part of the thermodynamic functions. Therefore, it makes sense to split the themodynamic functions as, e.g., the free energy, into three parts: the ideal term, the ring contribution and the terms beyond the ring contributions. The most important part beyond the ring contribution at lower temperatures provides the bound states, although this requires a completely different approach as, e.g., the quasi-chemical description \cite{EKK76,KKER86,EbFoFi17,EbFoFi17-b}. Here, we study high temperature expansions with respect to the interaction parameter $\xi_{ij} = - (\ell_{ij} / L_{ij})$, where $\ell_{ij}$ is the Coulomb length and $L_{ij}$ is another characteristic length for the short-range interactions. The parameter $\xi_{ij}$ is  assumed to be a small 
parameter in the region of interest. In the classical case, we identify the short-range length with the 
contact distance of the ions $L_{ij} = R_{ij}$ and in the quantum case with the relative thermal wave length
$L_{ij} = \lambda_{ij}$. We use here expansions with respect to $\xi_{ij}$ including the second and the 
third power and consider this for the systems of interest here (strong electrolytes and high-temperature plasmas) as an appropriate method for evaluating the free energy and the equation of state. At the same time, we try to include as many powers in $\kappa$ as possible.  
Density expansions for the pressure given in the literature for the classical case~\cite{Falkenhagen,Falkenhagen_b,
Yukhn80,EbFeKr21, EbFeKr21_b} as well as for the quantum case \cite{KKER86,AlMaMono} are only weakly convergent. For example, for the pressure  we get expansions of the form
\begin{eqnarray}
\beta p(n, T) = \sum_i n_i  - \frac{\kappa^3}{24 \piup} - 2 \piup \sum_{ij}  L_{ij} ^3 
\left[\frac{1}{6}  \ln (\kappa L_{ij}) + K_0(\xi_{ij}; s_a)  +\frac{1}{12} \right]+ O(n^{5/2} \ln n).
\label{pdensexp}
\end{eqnarray}
The functions $K_0(\xi)$  are known for the classical and for the quantum case  \cite{Falkenhagen,Falkenhagen_b,Friedman,Eb6768_a,Eb6768_b,Eb6768_c,Eb6768_d,Eb6768_e}.
By calculating the cluster expansions in detail and following the quantum case, we obtain in analogy to equation~\eqref{eq-19} the expression~\cite{Eb6768_a,Eb6768_b,Eb6768_c,Eb6768_d,Eb6768_e,EKK76,KKER86,AlMaMono}
\bea
K_0 (x) = \left[  - \frac{x}{6} + \frac{\sqrt{\piup}}{8} x^2 - \frac{1}{12} \left(C + 2 \ln 3 - \frac{1}{2}\right) x^3 + O(x^4) \right]  \pm 
({\rm s.e.}),
\label{eq-44}
\eea
where $({\rm s.e.})$ stands for symmetry effects which we neglect here, since they are small in the region of our interest. 
In the classical case, from the cluster expansion there follows the virial function  \cite{Falkenhagen,Falkenhagen_b,Friedman,Barthel}.
\bea 
K_0 (x) = - \frac{1}{2} - \frac{1}{2} x - \frac{1}{2} x^2 - \frac{1}{6} (C + \ln 3)  x^3 + O(x^4).
\eea
We see that the analogies between classical and quantum case are striking. 
A long experience shows that these density expansions including logarithmic terms are not appropriate for
practical calculations of the thermodynamic functions. 
What we expect, is that screening in combination with quantum effects leads to a kind of damping of the virial coefficients. This means that there is a tendency to decrease the virial coefficients, which leads to a better convergence of density virial series.

We show here that we need several steps to get representations of practical relevance. Our aim is to find 
improved virial expansions of the pressure of the form:
\bea
\beta p &=& \sum_i n_i  - \frac{\kappa^3}{24 \piup} \varphi (\kappa) - \frac{\piup}{3} \sum_{ij} n_i n_j \ell_{ij}^3 g ( \kappa) + \sum_{ij} n_i n_j B_{ij}'' (\kappa)  + \dots\,,
\label{eq-46}
\eea
where $\varphi (\kappa), g(\kappa), B_{ij}''(\kappa), \ldots$ are functions of the screening parameter $\kappa$ which are decreasing and  this way improve the convergence of the virial series in comparision to equation~\eqref{eq-44}.
The contributions to equation~\eqref{eq-46} have an increasing order in $e^2$, beginning with $e^3, e^4$, followed by 
$e^6,e^8$, etc.
In a different form, we may represent the cluster expansion of the pressure within the canonical ensemble in terms of densities \cite{KKER86,Alastuey95,Kahlbaum}.
In Debye--H\"uckel-like approximation we have $\varphi(x) = 1 - 3 \sqrt{\piup} x / 8 + (3/10) x^2 + \cdots$. We do not specify the screening functions $g(\kappa)$, and  $B(\kappa)$, which depend on the screening parameters $\kappa$ (canonical) and refer to \cite{KKER86}.  
A general expression for the contribution of the first ring diagrams is given in the Kelbg approximation
\cite{Kelbg63,Kelbg63-b,EbFoFi17,EbFoFi17-b}. Since the Kelbg potential in the original form and even in the exponential approximation given above has a complicated species dependence and leads to a matrix calculation, herein we used an averaging, leading  to the exponential potential with just one quantum parameter $\alpha$.
Note that  besides an averaged exponential parameter we may introduce an average effective interaction distance by $a = c/\alpha$. A problem is that different choices for the constant $c$ are in use as $c = 1$ or $c= 3/2$, depending on the reasons of convenient representations. Here, we use  $c=1$,
\bea
a = \frac{1}{\alpha} = \frac{\sqrt{\piup}}{4}
\frac{\sum_i \sum_j e_i^2 e_j^2 \lambda_{ij}}{\sum_i \sum_j e_i^2 e_j^2}.
\label{pairav2}
\eea
Note that using of an average effective quantum distance $a$ is sometimes convenient because of the similarity to the Debye theory.
At $T = 200$ kK, we estimate, e.g., for H-plasmas $a \simeq 0.40$~\AA~and for He-plasmas 
$a \simeq 0.26$~\AA.

In order to calculate the correlations in terms of these averaged parameters, we start from the Bogoliubov approximation for $g_{ij}$ given above, which has the same form in the quantum case with the only change 
that the $a$-parameter is now temperature-dependent. 
This approximation may be considered as the first term of a Bogoliubov-type expansion with respect to the plasma parameter for quantum plasmas~\cite{Falkenhagen,Falkenhagen_b,Yukhn80,EKK76}. 
Note that these roots are complex for larger $\kappa$ and this case needs
a special investigation (see section~\ref{sec-2} and~\cite{Holovko}). For very small densities, we find the mean Coulomb energy in ring approximation using the quantum length $a$:
\bea
U_{\rm C} =  \sum_i N_i u_i^{\rm C} = - \frac{V}{8 \piup }  \frac{\kappa^3}{\sqrt{1 + 2 \kappa a}} 
=  - \frac{V\kappa^3}{8 \piup} \bigg( 1 - \kappa a + \frac{3}{2} {\kappa^{2} a^2} - \frac{5}{2} {\kappa^{3} a^3}  - \dots\bigg).
\eea
The thermodynamic functions were already studied  above for the classical case. 
Since we brought the quantum potential to the same mathematical form, 
the function appearing in the quantum thermodynamics remains the same.

Our assumption is that these expansions are appropriate for quantum plasmas in the regions where no bound states exist and the overall interactions are weak, i.e., at $T > 10^5$ K.
The functions $\tau(\kappa),\varphi(\kappa)$, are a kind of extension of the limiting law in the spirit of the classical Debye--H\"uckel theory for charged hard spheres. The quantum form of the YK theory provides an improvement
of the quantum generalizations of the Debye theory, which was  studied earlier \cite{EKK76,EbFoFi17,EbFoFi17-b}.  
We derived for $k_{\rm B} T > I$ ($I$ is an ionization energy):
\bea
\varphi (\kappa) =  1 -  \frac{3 \sqrt{\piup}}{8} (\kappa \lambda) + \dots \,, \quad g(\kappa)  = \frac{\piup}{3} \ln(\gamma_2 \kappa \lambda).
\eea
Higher-order contributions to  $\varphi(x), \phi(x)$ were already studied in the context of the YK theory
 in~\cite{EbCMP25}. Within this scheme, we derived in section~\ref{sec-2.2} an approximation which now reads
 \bea
\varphi (\kappa) =  \frac{1}{\sqrt{1 +  (3 \sqrt{\piup}/4) \kappa \lambda}}.
\label{eq-50}
\eea
Note that the Coulomb energies of the ions and electrons found an important application 
in the ionization theory of plasmas \cite{RoLiEbRe25,EbRoAIP26}. The reason is as follows:
in dilute plasmas the ionization energy of an atom is given by the ground state energy 
$I = |E_{10}|$, which is the energy needed to bring an electron from the atom to the free states in a plasma. In a dense plasma, the ions and electrons do not have the energy zero but a negative self-energy corresponding 
to the potential hole formed by charges in a plasma. The ionization energy is lowered by an energy, which is in good approximation given by the Coulomb energy of the emitted electron plus the Coulomb energy of the 
ion getting free. 
This way we get for the ionization energy in YK approximation 
\bea
\Delta I = u_e^{\rm C} + u_i^{\rm C} = - \frac{(1 + Z^2)}{4 \piup \epsilon k_{\rm B} T}  \frac{e^2}{\sqrt{1 + 2 \kappa a}}, \quad a = \frac{3}{8} \sqrt{\piup} \lambda.
\eea
We mention here that for over 70 years, most calculations about ionization processes in star atmosheres, including extended tables have been performed using a variant of the ion sphere theory.
This is a theory leading to a cubic root depedence $\Delta I \sim n_i^{1/3}$ discussed 
already in the 2nd section which was developed in astrophysics by Uns\"old, Stewart, Pyatt and others \cite{Eb16,RoLiEbRe25}. It would be interesting to reconsider these calculations in the light of the 
Yukhnovskii--Kelbg theory which may permit to improve some of these astrophysical calculations.

\section{Including the ladder contribution}
\label{sec-4}
\begin{figure}[htbp]
\begin{center}
\includegraphics[scale=0.3]{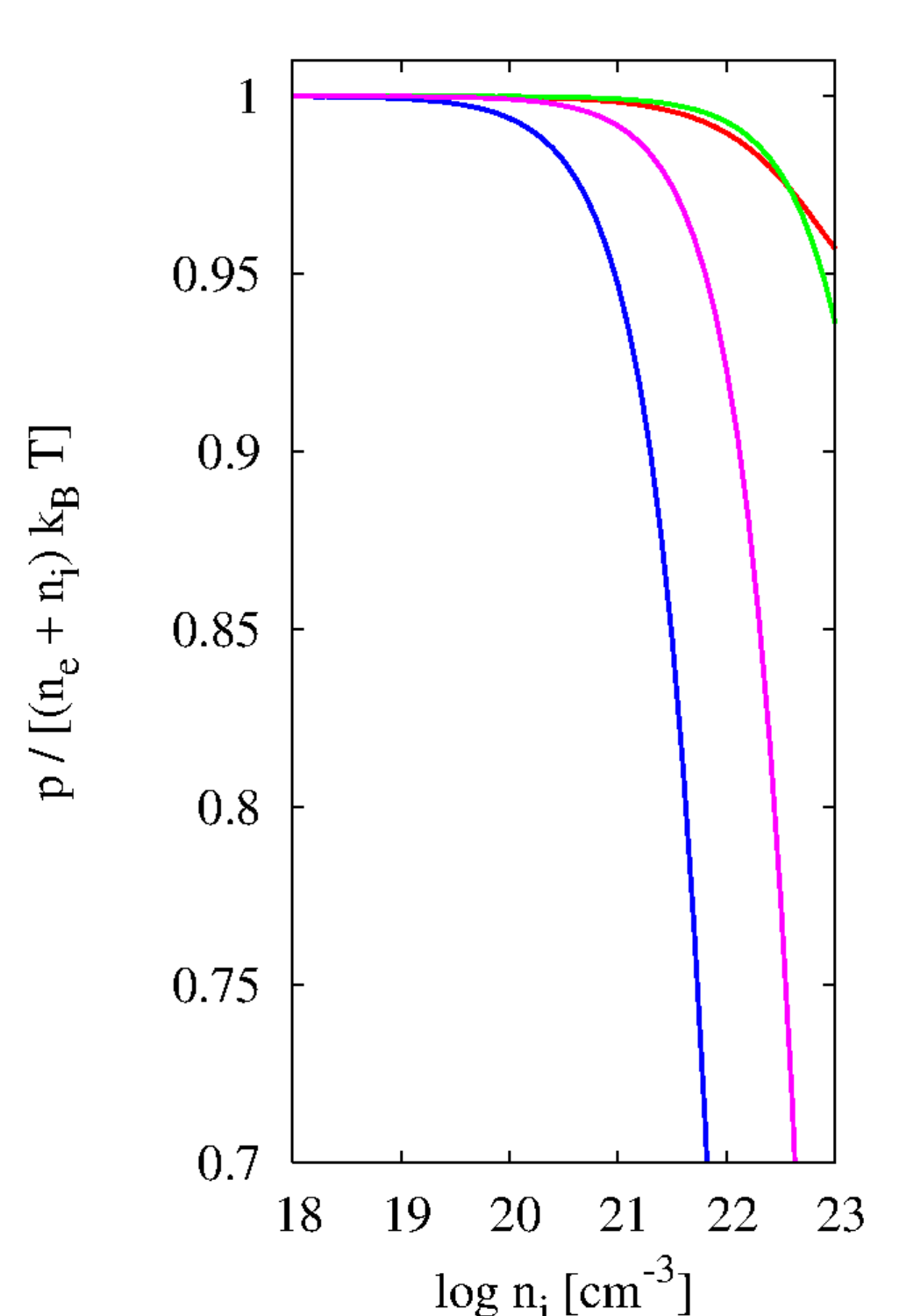}
\caption{(Colour online) The relative pressure for hydrogen and for helium plasmas  in Yukhnovskii--Kelbg approximation at
$T = 200$~kK. Comparison of the second order approximation (green for H and turquoise for He curves) with the third (logarithmic) approximation (red for H and blue for He). The curves for 2nd and 3rd approximation for hydrogen (green and red curves) are near  each other but those for Helium (blue and magenta lines) including the logarithmic term (blue curve) lead to larger effects due to  
the asymmetry of charges and masses. }
\label{fig-4}
\end{center}
\end{figure}

We study now the 3rd order contribution $O(e^6)$ to the pressure and other functions, which in graphical representation looks like a ladder with three rungs. For symmetrical systems like hydrogen plasmas, this
term nearly cancels out, up to minor effects due to mass asymmetry \cite{RoLiEbRe25}. For helium and other
complex plasmas, this contribution is of large relevance for the thermodynamics due to the asymmetry of the charges and masses. 
The 3rd order term provides a logarithmic contribution connected with the asymmetry of the charges and masses. This term is sometimes denoted as extended
limiting law~\cite{Falkenhagen,Falkenhagen_b,Friedman,KKER86}. The known exact low-density limit for the order $e^6$ reads for the free energy density and the pressure~\cite{KKER86}
\bea
\beta f_3 = - \frac{\piup}{3} \sum_{ij} n_i n_j \ell_{ij}^3 \bigg\{ \bigg[\ln (3 \kappa \lambda_{ij}) + \frac{C}{2}  - \frac{1}{2}\bigg] -
O (\kappa \lambda_{ij}) \dots\bigg\},\\
\beta p_3 = - \frac{\piup}{3} \sum_{ij} n_i n_j \ell_{ij}^3 \bigg\{\bigg[\ln (3 \kappa \lambda_{ij}) + \frac{C}{2} -  \frac{4}{3}\bigg] - O(\kappa \lambda_{ij}) \dots\bigg\}.
\eea
As mentioned, this contribution is of special importance for asymmetrical
systems like  solar plasmas including helium and lithium, where charges and masses are asymmetrical.
In order to get an orientation in what way to extend this to higher orders in $\kappa$, we may look again at the classical case (see section~\ref{sec-2} and~\cite{Falkenhagen,Falkenhagen_b,Friedman,EbKrCMP23,EbRoAIP26}).
Here, we use the more strict approach based on the Yukhnovskii--Kelbg theory. Based on the result for the Coulomb energy derived in section~\ref{sec-2.2} and using the virial theorem, in particular, equation~\eqref{eq-23}, for the 
third order contribution to the pressure we get 
\bea
\beta p_3 = -  \frac{\piup}{6} \sum_{ij} n_i n_j ( \ell_{ij} )^3 
  \big[\ln (2p_{ij}) + \ln (2s_{ij}) - 2 \ln (s_{ij} + p_{ij}) \big].
\eea
For $(\kappa \lambda) \rightarrow 0$, we get the correct limit
\bea
\beta p_3 \rightarrow  -  \frac{\piup}{3} \sum_{ij} n_i n_j ( \ell_{ij} )^3 \ln (\kappa \lambda_{ij}) + \dots \,.
\eea
The characteristic $\Gamma$-parameters are for typical plasmas:
\bea
\text{H:} \quad \Gamma = \Gamma_0, \qquad \text{He:} \quad \Gamma = 1.357 \Gamma_0, \quad \Gamma_0=\bigg( \frac{4\piup}{12 \piup \epsilon} n_e e^2  \bigg)^{1/3}.  
\eea
We also remember Kelbg's $a$-parameter which is a length proportional to $1/\sqrt{T}$. For example, this 
quantity is for hydrogen and helium given by
\bea
\text{H:} \quad a = \frac{\sqrt{\piup}}{16} (\lambda_{ee} + 2 \lambda_{ie} + \lambda_{ii}) , \qquad
\text{He:} \quad a = \frac{\sqrt{\piup}}{100} (\lambda_{ee} + 8 \lambda_{ie} + 16 \lambda_{ii}).
\eea
Summarizing, we get for the total pressure  in
YK-approximation the following closed expression valid for high-temperature plasmas
 \bea
\frac{p}{p_{\rm id}} = 1 - \frac{2 \piup }{3 n} \sum_{ij} n_i n_j \ell_{Ij}^2 \varphi(\kappa \lambda_{ij}) 
-\frac{\piup}{3 n} \sum_{ij} n_i n_j \ell_{ij}^3  g (\kappa \lambda_{ij}) + \dots \,.
\eea
We may check this result by comparison with available exact results for the EOS which were derived in the group of Kelbg in Rostock \cite{EbHoKe67,EbHoKe67-b}. According to these works, the exact results of quantum statistics for the functions $\varphi(x), g(x)$ read ($C$ is the Euler number)
\bea
\varphi(x)=  1 - (3/8) \sqrt{\piup} x + \dots \,, \quad g(x) = \ln(\gamma_2 x), \quad \ln \gamma_2 =\ln 3 + 2C - 4/3. 
\eea
Comparing our results obtained by means of the Kelbg--Yukhnovskii method we see that our function $\varphi(x)$ is in agreement with the exact result, but the function $g(x)$ differs by a constant. The reason is well-known:  the higher order contributions to the second order Bogoliubov expansion  equation~\eqref{eq-19} have an influence on the value of the constant. Taking into account these known effects, we get a corrected $g$-function
\bea
        g(x) = \ln \frac{\gamma_2 (\sqrt{\piup}/2) x}{\sqrt{1 + (\sqrt{\piup} /2) x}}, \qquad  \gamma_2 = 2.4960. 
\eea
For larger $\kappa$ leading to $x > 1/2$, we have to replace $g(x)$ (due to a different screening function, see section~\ref{sec-2.1}) and  assuming continuity, we get the expression:
\bea
          g(x) = \frac{\sqrt{2} \gamma_2}{4 x^2 -1} \ln \bigg(\frac{4 x}{2 x +1} \bigg), \quad \text{if} \quad x > 1/2. 
\eea
In a previous work \cite{EbRoAIP26} we provided a quantum-statistical derivation of the functions $\varphi(x)$
\bea
\varphi(x) = \frac{1 +(3/8) \sqrt{\piup} x}{1 + (3/4) \sqrt{\piup} x + (1/8) \sqrt{\piup} x^3},
\label{eq-62} 
\eea
which is consistent with our result and that obtained for $g(x)$, which includes higher orders in $\kappa \lambda$ and uses integral cosine and sine functions
\bea
g(\kappa \lambda_{ij}) = - \big[ \operatorname{Ci}(b_{ij} \kappa \lambda) + \operatorname{si}( b_{ij} \kappa \lambda) \big], \quad b_{ij} = 3 \lambda_{ij}/\sqrt{2 \piup}.
\label{eq-63} 
\eea 
Our $g$-function equations~\eqref{eq-62}--\eqref{eq-63} agrees with the quantum-statistical result at $x \rightarrow 0$ and is similar in shape.
However, we should say that the logarithmic contribution to the EOS still raises some open questions, we will work in figure~\ref{fig-4} with equation~\eqref{eq-50} and equation~\eqref{eq-62}.

Summarizing, we reached our aim to calculate the quantum-statistical contributions to the orders $e^4,e^6$ 
in the thermodynamic functions. Note that the combined action
of screening and quantum effects decreases the nonideality effects and leads to a  better convergence of the series and to a smoother behaviour of the equation of state. We pointed out in section~\ref{sec-2} that for electrolytes, the contributions of the  order $e^6$ were studied for
ionic systems with multiple charged ions \cite{Falkenhagen,Falkenhagen_b,Harned,Friedman,EbFeKr21, EbFeKr21_b,Holovko}. 
For the quantum case, the theory is in part still based on the analogies to electrolytes \cite{EbCMP25} and  detailed studies including the Yukhnovskii--Kelbg screening and the higher order $e^6$ were still missing. 
We have shown that the relatively simple Yukhnovskii--Kelbg theory provides in the second approximation developed here a curve (in blue in figure~\ref{fig-4}) which at small densities is in agreement with the quantum-statistical theory \cite{EbHoKe67,EbHoKe67-b,EKK76,EbFoFi17,EbFoFi17-b}.
We have demonstrated by using exponential potentials that the logarithmic effects are relevant not only for modelling the mass asymmetry in hydrogen \cite{RoLiEbRe25}, but, in particular, is 
important for helium and in the perspective also for sun plasmas due to the strong charge and mass asymmetries. 
We see much interest in applications to sun plasmas or other fusion plasmas in the  regions of full ionization.

\section{Conclusions}
\label{sec-4Concl}

We give an analysis of the theory of electrolytes and plasmas in equilibrium based on the use
of exponential potentials as it was first worked out already in the 1950th by Yukhnovskii and Kelbg.
We show that these tools are still very fruitful and flexible and may be generalized by including additional effects, such as short-range forces, quantum effects and high-density effects. The theory of exponential 
potentials provides for electrolytes an extension of the Debye--H\"uckel theory to higher densities.
For charged hard spheres this theory provides an alternative to the MSA, e.g., in the formulation of Henderson--Smith~\cite{EbKrCMP23,EbCMP25}. For this case, however, our theory does not provide improvements in accuracy in comparison with MSA. The situation may be  more advantageous for more realistic models for the ionic interaction in solutions with a soft core. In any case, this requires a transition from the simple averaged screening models used here to the  treatment of species-dependent exponential or more refined potentials such as the class of
Hellmann--Gurskii--Krasko pseudopotentials by means of a matrix theory  \cite{Sadykova}. The latter models also describe  potentials with 
easy Fourier transforms which permit to model the internal structure of more complex ions. 
For hydrogen-like quantum plasmas, already the simplest exponential potentials provide a good approximation 
to the known results of the quantum-statistical theory \cite{EbCMP25}.  We extended here the known results for the 
order  $e^4$ \cite{EbRoMDPI,Sadykova} to higher orders and calculated, in particular, the logarithmic 
contributions to the equation of state, which is of the order $e^6$. 

We pay special attention to electrolytes and plasmas in the regions where bound states do not play a role. Note that bound state contributions have the order $e^{2k}, k \geqslant 4$ \cite{EbCMP25}. We study further the role of the lower bound of the electrical energy found by Onsager and of cubic root effects at higher densities. In particular, we show  in this context, that the quadratic root laws, such as in the Debye--H\"uckel theory and related theories, may violate the Onsager bound. As far as we know, violations of Onsager's limit have never been observed so far in nature. Further, all the known Monte Carlo simulations for Coulomb systems observe Onsager's lower bound of the electrical energy \cite{HughForrest,DeWitt,BonitzPIMC_a,BonitzPIMC_b,BonitzPIMC_c}.
In this respect we could show here that the Yukhnovsky-Kelbg theory observes the 
Onsager bound for the Coulomb energy since the approximation leads
at higher concentrations/densities to a moderate increase as $n_i^{1/4}$ which is in accordance with
Onsager's bound.

We also discussed in this work  a structural transition of the pair distribution functions from 
exponential decay to damped oscillations. This transition is, however, not a thermodynamic phase transition
since the thermodynamic functions and their derivatives seem to be smooth.
The thermodynamic phase transitions in Coulomb systems, which were well studied in many 
works \cite{EKK76,EbFoFi17,EbFoFi17-b}, occur in the regions of densities and temperatures where bound states exist. 
Here we  study the phenomena in strong electrolytes without association effects and 
in high-temperature plasmas without bound states, referring for bound state effects to the other work 
\cite{EbFoFi17,EbFoFi17-b,EbRoAIP26}. In the region of full ionization, we observe only a structural transiion
to oscillations in the pair distribution. We have demonstrated that the relatively simple Yukhnovskii--Kelbg theory provides in the higher approximation developed here an equation of state (curve in blue in figure~\ref{fig-4}), which is in rather good accordance with the original quantum-statistical theory, which needs much higher efforts \cite{KKER86,EbFoFi17,EbFoFi17-b}.

\section*{Acknowledgement}
We express our thanks to many colleagues for suggestions, providing additional material and encouragement,
in particular to Gerd R\"opke from the Rostock University. Further we have benefited from remarks of referees
and express  sincere thanks to them.

\ukrainianpart

\title{До статистичної теорії сильних електролітів та високотемпературної плазми: нові застосування робіт Юхновського та Кельбга}
\author{В. Ебелінг\refaddr{label1},  М. Головко\refaddr{label2}}
\addresses{
	\addr{label1}{Інститут фізики, Університет Гумбольдта, Берлін, Німеччина},
	\addr{label2}{Інститут фізики конденсованих систем ім.~І.~Р.~Юхновськонго НАН України, вул. Свєнціцького, 1, 79011 Львів, Україна}
}

\makeukrtitle

\begin{abstract}
	\tolerance=3000%
	Згадуючи піонерські роботи Гюнтера Кельбга та Ігоря Юхновського зі статистичної фізики кулонівських систем, ми аналізуємо їхні методи та пропонуємо деякі нові застосування до іонних розчинів та квантової плазми. Зокрема, ми запропонували застосування теорії для сильних електролітів та для термічної високотемпературної плазми при $T > 10^5$~K з використанням моделі експоненційної взаємодії. Показано сильну структурну подібність цих двох класів кулонівських систем, поведінка яких визначається переважно внесками, пропорційними до $e^4$ та ~$e^6$. Передбачено, що при вищих густинах існує структурний перехід до немонотонних кореляцій. Термодинамічні функції демонструють плавний перехід від зростання, пропорційного квадратичному кореню, до повільнішого зростання типу $n_i^{1/4}$, що відповідає границі Онсагера. Вплив асиметрії зарядів та мас досліджується із застосуванням до багатосортних іонних систем та до високотемпературної плазми, зокрема до плазми з іонами He$^{2+}$.
	\keywords статистична фізика, сильні електроліти, високотемпературна плазма, покращена збіжність розвинень
	
\end{abstract}

\lastpage
\end{document}